\documentclass[twocolumn]{aastex701}

\usepackage{amsmath,amsfonts,amssymb}
\usepackage{bm}
\usepackage{graphics}
\usepackage{graphicx}
\usepackage{hyperref}
\usepackage{here} 
\usepackage{type1cm}
\usepackage{multirow}
\usepackage{float}
\usepackage{color}
\usepackage{ulem}

\hypersetup{
colorlinks=true,
linkcolor=blue,
citecolor=blue,
urlcolor=blue
}

\def\kms{\hbox{km s$^{-1}$}}
\def\VLSR{\hbox{$V_{\rm LSR}$}}

\def\sun{\hbox{$\odot$}}
\def\earth{\hbox{$\oplus$}}
\def\lesssim{\mathrel{\hbox{\rlap{\hbox{\lower4pt\hbox{$\sim$}}}\hbox{$<$}}}}
\def\gtrsim{\mathrel{\hbox{\rlap{\hbox{\lower4pt\hbox{$\sim$}}}\hbox{$>$}}}}

\def\arcdeg{\hbox{$^\circ$}}

\def\Hone{\hbox{H\,{\scriptsize I}}}

\shorttitle{Signature of a Dark Ultra-compact Dwarf Galaxy}
\shortauthors{Udagawa et al.}

\begin{document}

\title{Signature of a Dark Ultra-compact Dwarf Galaxy Transiting the Milky Way Disk}

\correspondingauthor{Tomoharu Oka}
\email{tomo@phys.keio.ac.jp}

\author[gname=Ken, sname=Udagawa]{Ken Udagawa}
\affiliation{School of Fundamental Science and Technology, Graduate School of Science and Technology, Keio University, 3-14-1 Hiyoshi, Kohoku-ku, Yokohama, Kanagawa 223-8522, Japan} 
\email{udg-wkn.0806@keio.jp}

\author[orcid=0000-0002-5566-0634,gname=Tomoharu,sname=Oka]{Tomoharu Oka} 
\affiliation{School of Fundamental Science and Technology, Graduate School of Science and Technology, Keio University, 3-14-1 Hiyoshi, Kohoku-ku, Yokohama, Kanagawa 223-8522, Japan}
\affiliation{Department of Physics, Institute of Science and Technology, Keio University, 3-14-1 Hiyoshi, Kohoku-ku, Yokohama, Kanagawa 223-8522, Japan}
\email{tomo@phys.keio.ac.jp}

\author[orcid=0000-0003-3853-1686, gname=Hiroki, sname=Yokozuka]{Hiroki Yokozuka}
\affiliation{School of Fundamental Science and Technology, Graduate School of Science and Technology, Keio University, 3-14-1 Hiyoshi, Kohoku-ku, Yokohama, Kanagawa 223-8522, Japan} 
\email{gackt4869akai@keio.jp}

\author[orcid=0009-0006-9842-4830,gname=Tatsuya,sname=Kotani]{Tatsuya Kotani}
\affiliation{School of Fundamental Science and Technology, Graduate School of Science and Technology, Keio University, 3-14-1 Hiyoshi, Kohoku-ku, Yokohama, Kanagawa 223-8522, Japan} 
\email{sci.tatsu.729@keio.jp}

\begin{abstract}
We report the discovery of a vertical velocity anomaly (VVA) in the stellar component of the Galactic disk, consistent with the impact of a dark, ultra-compact dwarf galaxy (UCD)-sized object plunging into the Milky Way. The anomaly spatially coincides with a suite of gaseous disturbances---including an \Hone\ void, a molecular shell (CO 16.134--0.553), and a vertical \Hone\ filament---previously interpreted as signatures of a dark matter subhalo (DMSH) collision. Analysis of Gaia DR2 astrometry reveals a statistically significant vertical velocity dip co-located with these features, supporting a dynamical origin. The absence of a luminous source at the filament's terminus suggests a dark or failed UCD. These results provide rare observational evidence for a low-mass DMSH, with implications for the substructure of dark matter halos and constraints on the $\Lambda$CDM model. 
\end{abstract}


\section{Introduction} \label{sec:intro}
Dark matter (DM) constitutes approximately 26\,\% of the universe's mass and exceeds the mass of visible baryonic matter by more than a factor of five \citep{Planck20}. It forms extended halos around galaxies and clusters, playing a critical role in cosmic structure formation. The leading candidate for DM is a population of cold, slow-moving, weakly interacting massive particles (WIMPs) \citep{deSwart17}. The cold dark matter model with a cosmological constant ($\Lambda$CDM) has successfully explained many large-scale phenomena, such as the cosmic microwave background anisotropies, galaxy clustering, primordial element abundances, and cosmic acceleration \citep{Riess98,Perlmutter99,Spergel03,Tegmark04}.

Nonetheless, the $\Lambda$CDM paradigm faces notable challenges on smaller scales \citep{DelPopolo17}. Chief among these is the ``missing satellite problem," wherein $N$-body simulations predict an order of magnitude more dark matter subhalos (DMSHs) around Milky Way-like galaxies ($M_{\rm halo} \sim 10^{12}\, M_\odot$) than the observed population of luminous satellite galaxies \citep{Klypin99,Diemand07}. While discoveries of ultra-faint dwarf galaxies have partially alleviated this discrepancy \citep{Willman05,Belokurov06,Zucker06,Sakamoto06,Irwin07}, the population of low-mass ($\lesssim 10^8\, M_\odot$) DMSHs remains largely unconstrained due to their lack of baryonic tracers and the resolution limits of simulations.

\begin{figure*}[htbp] 
\centering
\includegraphics[width=\linewidth]{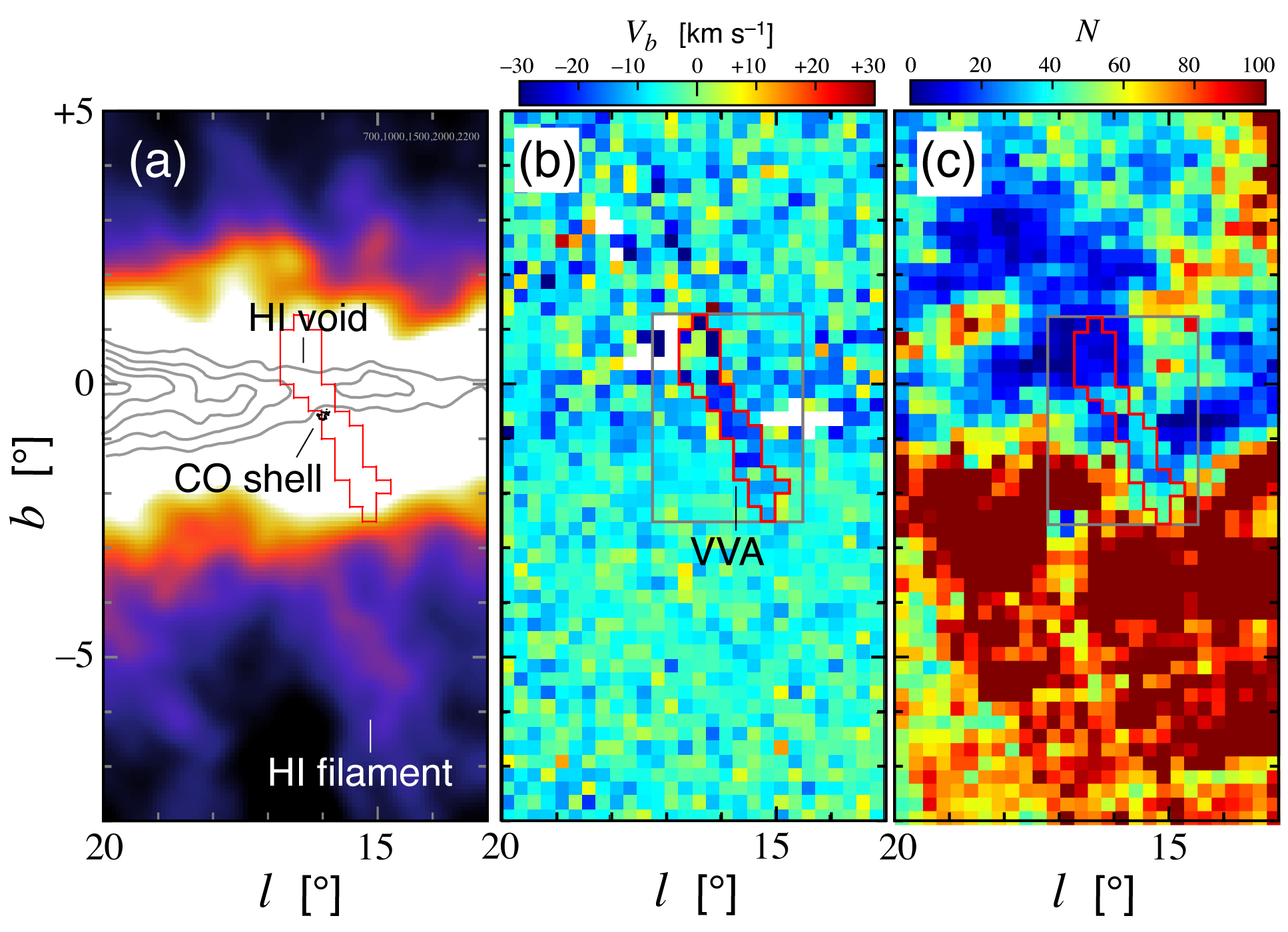}
\caption{Discovery of the vertical velocity anomaly. (a) Schematic view of the anomalous features in gaseous components.  The color map shows \Hone\ emission integrated over velocities between $\VLSR\! =\! 40\,\kms$ and $120\,\kms$. Gray contours show \Hone\ emission integrated over velocities between $\VLSR\! =\! 60\,\kms$ and $100\,\kms$ at levels of $700, 1000, 1500, 2000$, and $2200$ K \kms .  (b) Distribution of the mean latitudinal (vertical) velocity of stars ($V_b$) at the distance of $3.9\mbox{--}4.0$ kpc.  Pixels which contain fewer than 5 stars are shown in white.  The area enclosed by solid red lines defines the vertical velocity anomaly (VVA). }
\label{fig1}
\end{figure*}

A promising candidate for a small DMSH was identified as a broad-velocity-width molecular feature (BVF), CO 16.134--0.553, discovered during a systematic CO {\it J}=1--0 survey of the Galactic disk with $\sim 20^{\prime\prime}$ resolution \citep{Yokozuka21,Umemoto17}. This object spans $\sim$5\,pc, exhibits a line width of $\sim\!30\,\kms$, and lies at a distance of $\sim\!4\,\mbox{kpc}$ from the Sun. It shows no evident driving source, yet displays strong SiO emission indicative of dissociative shocks \citep{Yokozuka24}, suggesting a dynamic origin. CO 16.134--0.553 marks the eastern boundary of a $\sim$15\,pc CO shell, spatially coincident with a 1$^\circ$-diameter \Hone\ void and a vertically oriented, 4$^\circ$-long \Hone\ filament (Figure~\ref{fig1}a).

The alignment of the CO shell, \Hone\ void, and filament suggests a high-velocity impact by a massive compact object, possibly a small DMSH containing some baryonic matter \citep{Yokozuka24}. This scenario is supported by hydrodynamic simulations that reproduce the observed gaseous structures \citep{Bekki06,Tepper-Garcia18,Shah19}. Estimated parameters for the intruder---size 15--70\,pc, velocity $\sim$130\,km\,s$^{-1}$, and mass $\sim 6 \times 10^7\,M_\odot$---place it within the expected range for low-mass DMSHs \citep{Bovy17,Garrison17}. However, these estimates are derived solely from gas morphology and simple geometry. Independent evidence, especially from stellar kinematics, is crucial to validate the DMSH impact scenario and constrain its physical properties.

\section{Data and Analysis} \label{sec:data}
An event involving the plunge of a DMSH is expected to leave discernible traces within both the stellar population and the gaseous components of the Galactic disc. Motivated by this, we utilized astrometric data from the second data release of Gaia \citep[Gaia DR2;][]{Gaia16, Gaia18}, which includes 1,015,747 sources within our area of interest: $+13{\overset{\circ}{.}}0\!\leq\! l \!\leq\! +20{\overset{\circ}{.}}0$, $-8{\overset{\circ}{.}}0 \!\leq\! b \!\leq\! +5{\overset{\circ}{.}}0$, and $3.5\,\mathrm{kpc} \!\leq\! D \!\leq\! 4.5\,\mathrm{kpc}$. 

We converted the positions and proper motions of stars from equatorial (ICRS) coordinates to Galactic coordinates using the transformation procedure outlined in the Gaia Data Release Documentation provided on the ESA website\footnote{\url{https://gea.esac.esa.int/archive/documentation/GDR2/Data_processing/chap_cu3ast/sec_cu3ast_intro/ssec_cu3ast_intro_tansforms.html\#SSS1}}. 

Subsequently, we subtracted for the systemic velocity due to Galactic rotation from the longitudinal velocities. Employing the axisymmetric, circularly rotating thin-disk model for the Milky Way Galaxy, the transverse velocity of an observed object relative to the local reference frame of stars in the solar neighborhood (the Local Standard of Rest; LSR) is given by the equation, 
\begin{equation}
  \label{galactic_rotation_l}
  V_t=\left(\frac{\Theta}{R}-\frac{\Theta_0}{R_0}\right)R_0\cos{l}-\frac{\Theta}{R}D_p .
\end{equation}
where $D_p=(1/\varpi)\cos{b}$ is the object's distance from the sun projected onto the Galactic plane, $\varpi$ is the parallax in mas,  $R$ is that from the Galactic center (Galactic radius), 
$\Theta(R)$ is the rotation velocity at $R$, $R_0$ is the Galactic radius of the sun, and $\Theta_0\!\equiv\! \Theta(R_0)$.  The Galactic rotation velocity is modeled using a power-law: $\Theta(R) = \Theta_0(R/R_0)^\alpha$, with $(R_0,\Theta_0, \alpha)=(8.27\,\mathrm{kpc}, 248\,\mathrm{km\,s^{-1}}, 0.05)$ \citep{Honma12}.  Since the moving direction along the Galactic rotation on the celestial sphere is parallel to the Galactic longitudinal axis, we subtracted the transverse velocity from the longitudinal velocities, 
\begin{equation}
  V_{l,c}=V_l-V_t .
\end{equation}
In our analyses, we used $V_{l,c}$ and $V_b$ of stars to calculate the longitudinal and latitudinal velocity distributions.  

We then averaged the rotation-corrected proper motions within each $0{\overset{\circ}{.}}25 \times 0{\overset{\circ}{.}}25 \times 0.1\,\mathrm{kpc}$ voxel to depict the spatial distribution of stellar velocities.

\begin{figure}[htbp] 
\centering
\includegraphics[width=\linewidth]{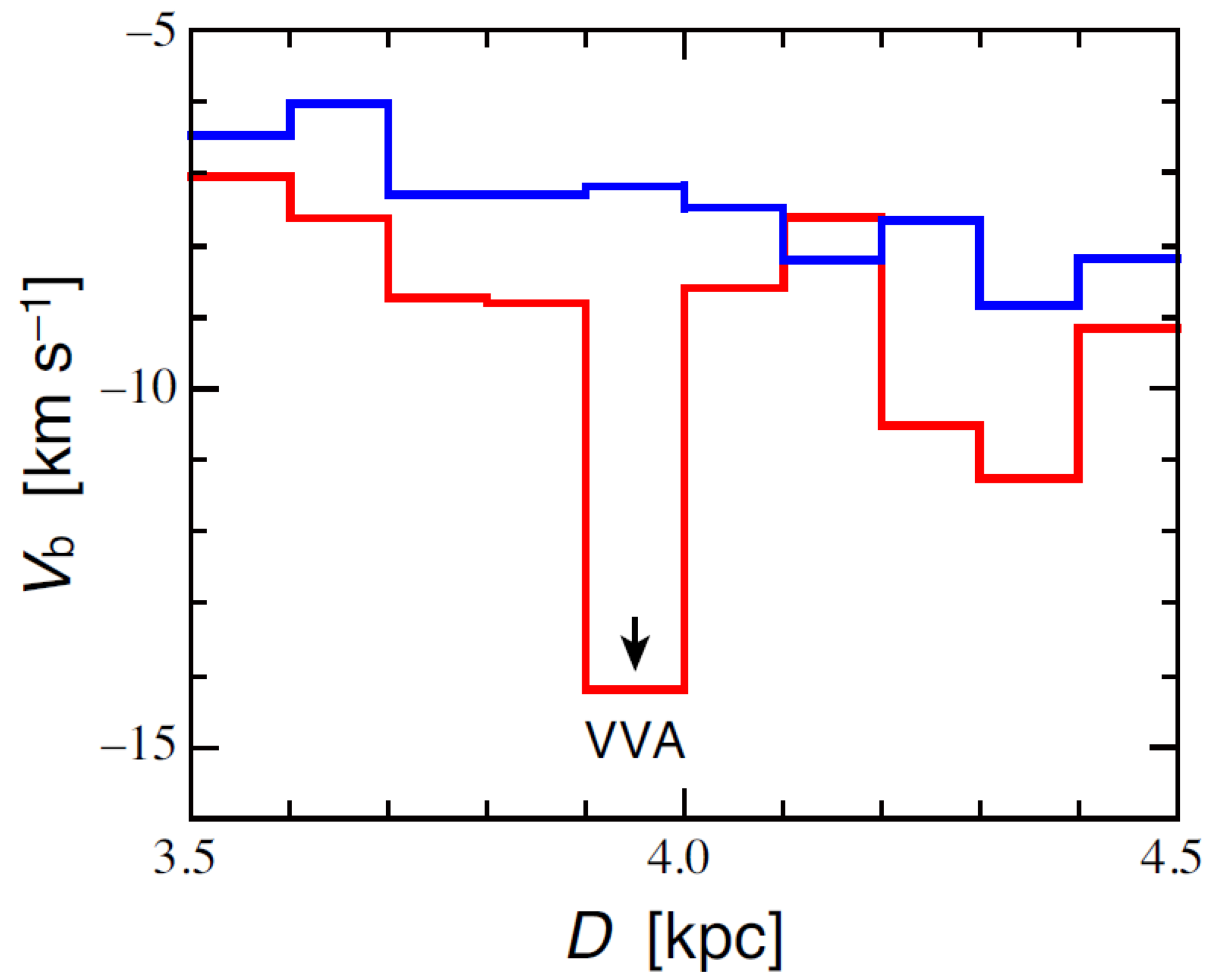}
\caption{Plot of the average vertical velocity vs. distance. The red line shows the average vertical velocity of stars in the VVA, and the blue line shows that in the surrounding area (inside the gray rectangle in Figure \ref{fig1}(b) excluding the VVA).}
\label{fig2}
\end{figure}

\section{Results} \label{sec:results}
The analysis revealed a vertical velocity anomaly (VVA) at the root of the \Hone\ filament, as shown in Figure \ref{fig1}(b), with coordinates $(l,b,D)\!\simeq\!(15{\overset{\circ}{.}}8, -0{\overset{\circ}{.}}5, 3.9\mbox{--}4.0\,\mathrm{kpc})$. We defined the spatial extent of this VVA with solid red lines in Figure \ref{fig1}. Within this anomaly, the average vertical (latitudinal) velocity of stars is typically $7\,\mathrm{km\,s^{-1}}$ lower than in the surrounding areas or at adjacent distances, as illustrated in Figure \ref{fig2}. 

\begin{figure}[htbp] 
\centering
\includegraphics[width=\linewidth]{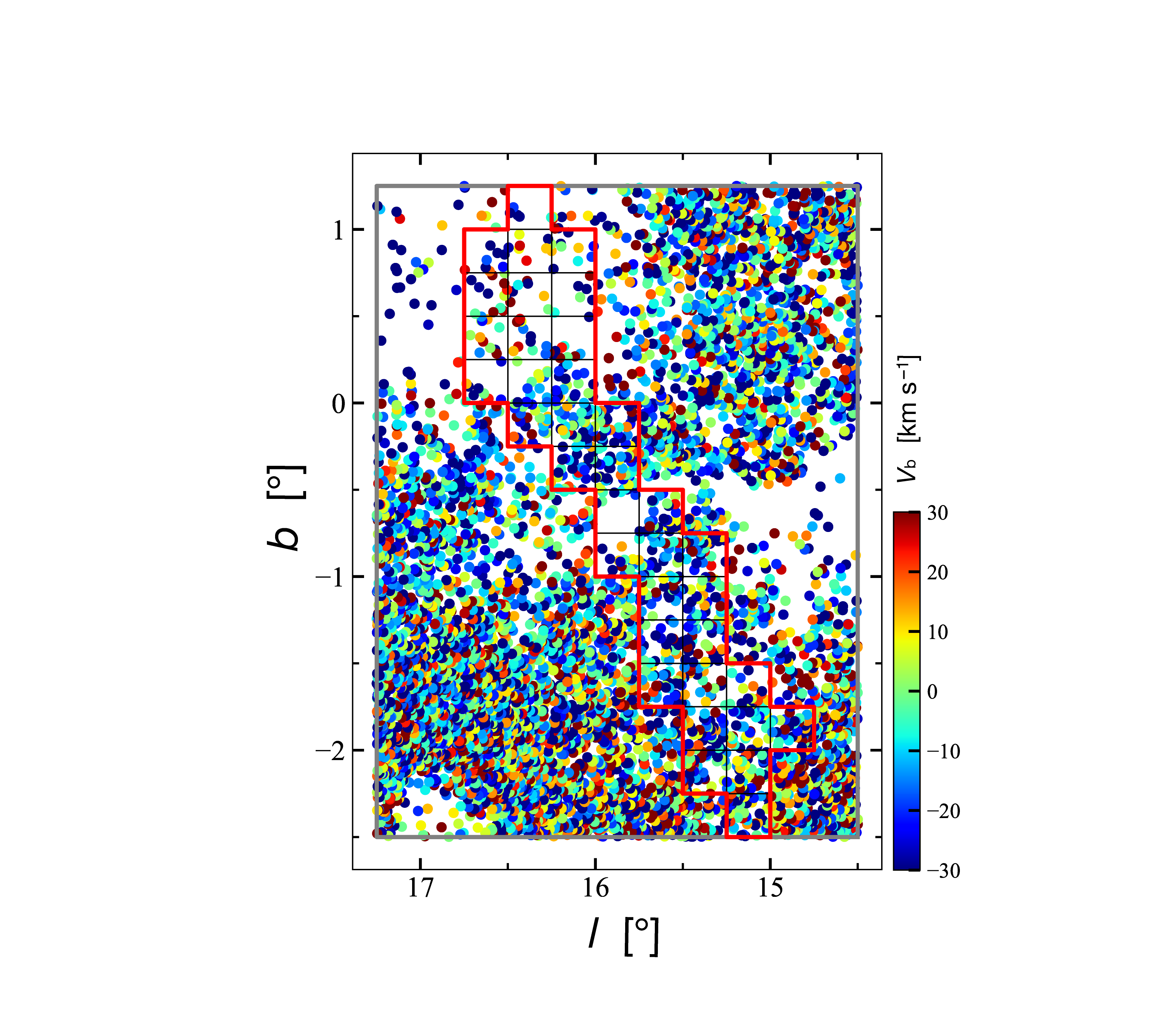}
\caption{Spatial distribution of individual Gaia DR2 stars within the rectangular region indicated in Figure \ref{fig1}b. Each dot represents a single star and is color-coded by its vertical velocity ($V_b$). The velocity scale is identical to that used in Figure 1b. A coherent group of stars with systematically negative vertical velocities is apparent, forming the vertical velocity anomaly (VVA) at the center of the frame. This visualization complements the binned median velocity map in Figure \ref{fig1}b by showing the unbinned stellar distribution underlying the VVA structure.}
\label{fig3}
\end{figure}

Although the VVA manifests as a statistically significant dip in the mean vertical velocity field, the distribution of individual stellar velocities exhibits substantial dispersion. Visualization of individual stars (Figure \ref{fig3}) reveals no obvious coherent concentration of stars with uniformly negative velocities. This suggests that the anomaly reflects a shift in the statistical mean of a broadly scattered velocity distribution, rather than a spatially clustered population with coherent motion.

A $z$-test confirmed that this deviation in average vertical velocity is statistically significant, reaching a $6\sigma$ level. No notable variations were observed in the horizontal (longitudinal) velocity distribution, indicating that stars within the VVA predominantly exhibit negative vertical velocities. The VVA is elongated, measuring approximately 200 pc in length and 50 pc in width. Its three-dimensional location coincides with the center of the \Hone\ void at $(l,b) \simeq (16{\overset{\circ}{.}}2, -0{\overset{\circ}{.}}05)$ and overlaps significantly with the CO shell at $(l,b) \simeq (16{\overset{\circ}{.}}05, -0{\overset{\circ}{.}}55)$, reinforcing the physical link between the VVA and the previously identified gaseous anomalies. 

\section{Discussion} \label{sec:discussion}
The width of the VVA, approximately 50 pc, may offer a preliminary estimate for the size of the hypothesised DMSH. The lower limit of the DMSH's mass can be derived from the conditions for tidal stability, expressed as ${GM_\mathrm{DMSH}}/{r_\mathrm{DMSH}^2} \gtrsim {GM_\mathrm{gal} r_\mathrm{DMSH}}/{R^3}$, where $R$ is the Galactocentric distance and $ r_\mathrm{DMSH}$ is the radius of the DMSH. Assuming $M_\mathrm{gal} \simeq 4\times 10^{10}\,M_\odot$ \citep{Valenti16}, $R=4.6\,\mathrm{kpc}$ and $r_\mathrm{DMSH}=25\,\mathrm{pc}$, the estimated lower limit for the DMSH mass is $M_\mathrm{DMSH} \gtrsim 6\times 10^3\,M_\odot$. Another method for estimating mass is via the Hoyle-Lyttleton radius, $r_\mathrm{HL}=2GM_\mathrm{DMSH}/V_\mathrm{pl}^2$, where $V_\mathrm{pl}$ is the plunging velocity. Using $V_\mathrm{pl}=130\, \mathrm{km\,s^{-1}}$ \citep{Yokozuka24}  and assuming $r_\mathrm{HL}\simeq r_\mathrm{DMSH}=25\,\mathrm{pc}$, the calculated mass of the DMSH is approximately $M_\mathrm{DMSH}\!\simeq\! 5\times 10^7\,M_\odot$.

\begin{figure*}[htbp] 
    \centering
    \includegraphics[width=\linewidth]{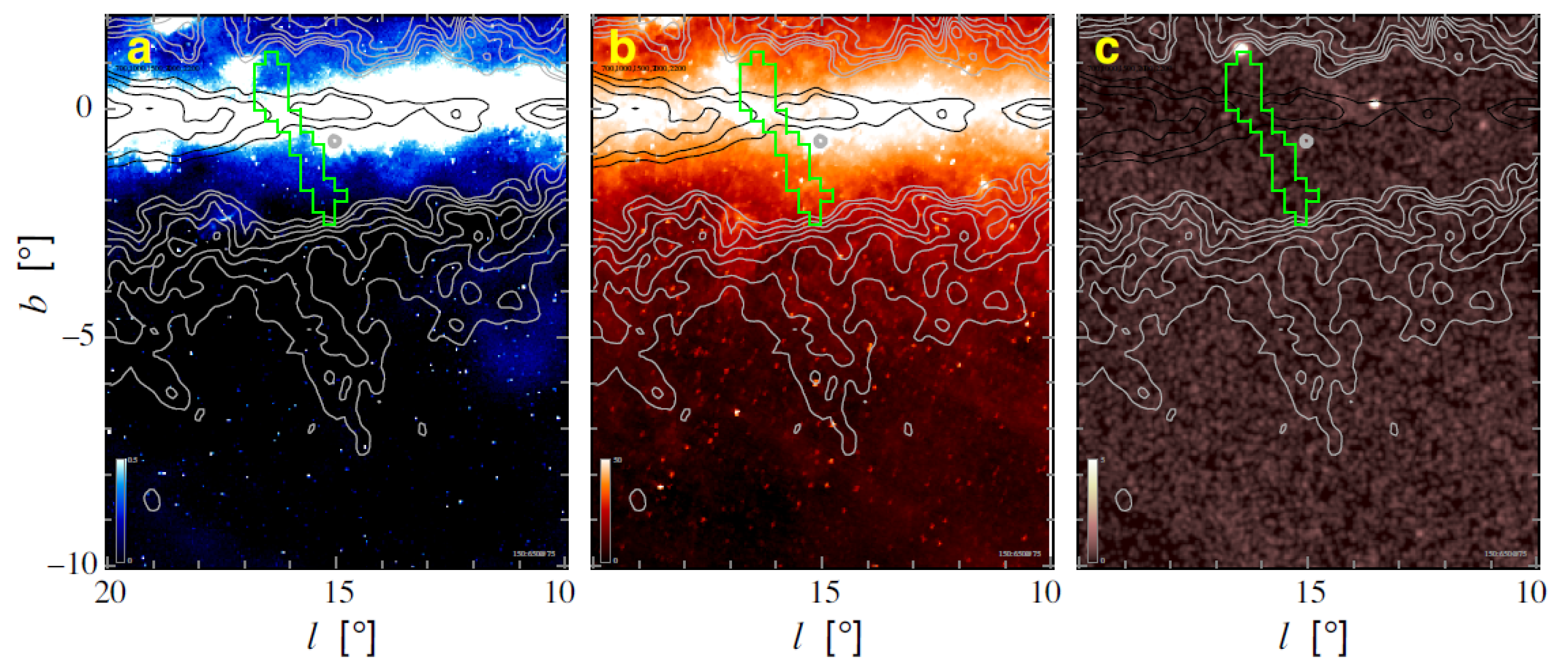}
    \caption{Multiwavelength views (radio, infrared, X-ray) of the region surrounding the \Hone\ filament, highlighting the absence of luminous counterparts at its tip.  Colour images of (a) 170--231 MHz radio continuum, (b) IRAS 12 $\mu$m, and (c) 0.1--2.4 keV X-ray with contours of \Hone\ integrated intensity.  Light gray contours show \Hone\ emission integrated over velocities between $\VLSR\! =\! 40\,\kms$ and $120\,\kms$ at levels from 150  K \kms\ with 75  K \kms\ intervals.  Dark gray contours show \Hone\ emission integrated over velocities between $\VLSR\! =\! 60\,\kms$ and $100\,\kms$ at levels of $700, 1000, 1500, 2000$, and $2200$ K \kms .  The VVA is shown by solid green lines.}
    \label{fig4}
\end{figure*}

Unlike global perturbations caused by massive intruders such as the Sagittarius dwarf, which produce disk-wide warps and corrugations \citep[e.g.,][]{Bland-Hawthorn21}, the features discussed here are highly localized. This is consistent with a lower-mass transiting object that does not significantly displace the disk center-of-mass. The absence of a significant recoil or global disk response implies that the impactor's mass is moderate, consistent with our DMSH-scale estimate.  The possible presence of large outer \Hone\ holes in galaxies \citep[e.g.,][]{Bosma17} raises intriguing parallels, although such features may originate from different mechanisms including stellar feedback.

The multiwavelength context of the \Hone\ filament is shown in Figure \ref{fig4}.  The absence of any luminous sources across radio, infrared, and X-ray bands at the tip of the \Hone\ filament suggests that the DMSH responsible for the formation of the VVA might represent a `dark' or `failed' dwarf galaxy, with a mass below $10^9\,M_\odot$. A notable comparison is Smith's Cloud \citep{Smith63}, a high-velocity \Hone\ cloud that crossed the Galactic disk approximately 70 million years ago. To have endured this passage, Smith's Cloud is likely encapsulated within a DMSH exceeding $10^8\, M_\odot$ \citep{Nichols09}. Our findings suggest a smaller analog for such dark dwarf galaxies, albeit considerably smaller than the former, with dimensions akin to those of ultra-compact dwarfs (UCDs) \citep{Drinkwater00}, typically found in dense galaxy clusters.  UCDs are characterized by old stellar populations, half-light radii of $15\mbox{--}100$ pc, and masses of $10^{6.5\mbox{--}8}\,M_\odot$ \citep{Mieske08}. The possible existence of a `failed' UCD---devoid of any detectable stellar component---stands in direct conflict with the prevailing formation scenario, which holds that UCDs are the stripped remnants of dwarf galaxies. If confirmed, such an object would fundamentally challenge our current understanding of compact stellar systems and their dark matter environments.  Further insights into the structure of the VVA may be gained through spatial statistical analysis, such as the methods outlined by \citet{Baddeley15}.

\section{Summary} \label{sec:summary}
The discovery of the vertical velocity anomaly (VVA) provides compelling evidence that a low-mass dark matter subhalo (DMSH) has plunged into the Milky Way's disk, generating a coherent set of gaseous anomalies. This finding not only supports the feasibility of using stellar kinematics to detect otherwise invisible substructures, but also motivates systematic searches for similar signatures---gaseous filaments, voids, shells, BVFs, and stellar VVAs---throughout the Galactic disk. The object inferred here, potentially a `failed' ultra-compact dwarf galaxy, exemplifies a previously unrecognized class of dark halo structures. Its detection opens a new observational window onto the low-mass end of the DMSH population, offering a promising avenue to critically test the small-scale predictions of the standard $\Lambda$CDM cosmology and deepen our understanding of galaxy formation and dark matter.  

Future surveys targeting similar kinematic and gaseous anomalies across the Galactic disk will be essential to assess the frequency and diversity of such dark subhalo events.  Detailed study of the stellar populations associated with the VVA---including identification of likely member stars---may offer important clues to the nature of the impactor.  Statistical modeling of the stellar distribution, such as mixture-based analysis or spatial point process methods, may further constrain the structure and origin of the velocity anomaly. These approaches will be pursued in future work.

\begin{acknowledgments}
This work made use of data from the European Space Agency (ESA) mission \textit{Gaia} (\url{https://www.cosmos.esa.int/gaia}), processed by the \textit{Gaia} Data Processing and Analysis Consortium (DPAC, \url{https://www.cosmos.esa.int/web/gaia/dpac/consortium}). Funding for DPAC has been provided by national institutions, in particular those participating in the \textit{Gaia} Multilateral Agreement. We thank Editage (\url{www.editage.jp}) for English language editing. T.O. acknowledges support from JSPS KAKENHI Grant Number 20H00178.
\end{acknowledgments}


\facilities{Gaia, NRO 45m Telescope (FUGIN), HI4PI Survey}

\software{MATLAB}


\begin{thebibliography}{99}
\bibitem[Baddeley et al.(2015)]{Baddeley15} Baddeley, A., Rubak, E., \& Turner, R.\ 2015, \textit{Spatial Point Patterns: Methodology and Applications with R} (Boca Raton, FL: CRC Press)
\bibitem[Bekki \& Chiba(2006)]{Bekki06} Bekki, K., \& Chiba, M. 2006, ApJL, 637, L97
\bibitem[Belokurov et al.(2006)]{Belokurov06} Belokurov, V., Zucker, D. B., Evans, N. W., et al. 2006, ApJL, 647, L111
\bibitem[Bland-Hawthorn \& Tepper-Garc\'ia(2021)]{Bland-Hawthorn21} Bland-Hawthorn, J., \&\ Tepper-Garc\'ia, T.  2021, NMRAS, 504, 3168 
\bibitem[Bosma(2017)]{Bosma17} Bosma, A.  2017, Astrophys. Space Sci. Lib., 434, 209
\bibitem[Bovy et al.(2017)]{Bovy17} Bovy, J., Erkal, D., \& Sanders, J. L. 2017, MNRAS, 466, 628
\bibitem[Bullock \& Boylan-Kolchin(2017)]{Bullock17} Bullock, J. S., \& Boylan-Kolchin, M. 2017, ARA\&A, 55, 343
\bibitem[de Swart et al.(2017)]{deSwart17} de Swart, J., Bertone, G., \& van Dongen, J. 2017, NatAs, 1, 59
\bibitem[Del Popolo \& Le Delliou(2017)]{DelPopolo17} Del Popolo, A., \& Le Delliou, M. 2017, Galaxies, 5, 17
\bibitem[Diemand et al.(2007)]{Diemand07} Diemand, J., Kuhlen, M., \& Madau, P. 2007, ApJ, 657, 262
\bibitem[Drinkwater et al.(2000)]{Drinkwater00} Drinkwater, M. J., Jones, J. B., Gregg, M. D., et al. 2000, PASJ, 17, 227
\bibitem[Gaia Collaboration et al.(2016)]{Gaia16} Gaia Collaboration, Prusti, T., de Bruijne, J. H. J., et al. 2016, A\&A, 595, A1
\bibitem[Gaia Collaboration et al.(2018)]{Gaia18} Gaia Collaboration, Brown, A. G. A., Vallenari, A., et al. 2018, A\&A, 616, A1
\bibitem[Garrison-Kimmel et al.(2017)]{Garrison17} Garrison-Kimmel, S., Wetzel, A., Bullock, J. S., et al. 2017, MNRAS, 471, 1709
\bibitem[Honma et al.(2012)]{Honma12} Honma, M., Nagayama, T., Ando, K., et al. 2012, PASJ, 64, 136
\bibitem[Irwin et al.(2007)]{Irwin07} Irwin, M. J., Belokurov, V., Evans, N. W., et al. 2007, ApJL, 656, L13
\bibitem[Klypin et al.(1999)]{Klypin99} Klypin, A. A., Kravtsov, A. V., Valenzuela, O., \& Prada, F. 1999, ApJ, 522, 82
\bibitem[Mieske \& Kroupa(2008)]{Mieske08} Mieske, S., \& Kroupa, P. 2008, ApJ, 677, 276
\bibitem[Nichols \& Bland-Hawthorn(2009)]{Nichols09} Nichols, M., \& Bland-Hawthorn, J. 2009, ApJ, 707, 1642
\bibitem[Perlmutter et al.(1999)]{Perlmutter99} Perlmutter, S., Aldering, G., Goldhaber, G., et al. 1999, ApJ, 517, 565
\bibitem[Planck Collaboration et al. (2020)]{Planck20} Planck Collaboration, Aghanim, N., Akrami, Y., et al. 2020, A\&A, 641, A6
\bibitem[Riess et al.(1998)]{Riess98} Riess, A. G., Filippenko, A. V., Challis, P., et al. 1998, AJ, 116, 1009
\bibitem[Sakamoto \& Hasegawa(2006)]{Sakamoto06} Sakamoto, T., \& Hasegawa, T. 2006, ApJL, 653, L29
\bibitem[Shah et al.(2019)]{Shah19} Shah, M., Bekki, K., Vinsen, K., \& Foster, S. 2019, MNRAS, 482, 4188
\bibitem[Smith(1963)]{Smith63} Smith, G. P. 1963, Bull. Astron. Inst. Neth., 17, 203
\bibitem[Spergel et al.(2003)]{Spergel03} Spergel, D. N., Verde, L., Peiris, H. V., et al. 2003, ApJS, 148, 175
\bibitem[Tegmark et al.(2004)]{Tegmark04} Tegmark, M., Strauss, M. A., Blanton, M. R., et al. 2004, PhRvD, 69, 103501
\bibitem[Tepper-Garc\'ia \& Bland-Hawthorn(2018)]{Tepper-Garcia18} Tepper-Garc\'ia, T., \& Bland-Hawthorn, J. 2018, MNRAS, 473, 5514
\bibitem[Umemoto et al.(2017)]{Umemoto17} Umemoto, T., Minamidani, T., Kuno, N., et al. 2017, PASJ, 69, 78
\bibitem[Valenti et al.(2016)]{Valenti16} Valenti, E., Zoccali, M., Gonzalez, O. A., et al. 2016, A\&A, 587, L6
\bibitem[Willman et al.(2005)]{Willman05} Willman, B., Dalcanton, J. J., Martinez-Delgado, D., et al. 2005, ApJL, 626, L85
\bibitem[Yokozuka et al.(2021)]{Yokozuka21} Yokozuka, H., Oka, T., Takekawa, S., Iwata, Y., \& Tsujimoto, S. 2021, ApJ, 908, 246
\bibitem[Yokozuka et al.(2024)]{Yokozuka24} Yokozuka, H., Oka, T., Tsujimoto, S., Iwata, Y., \& Kaneko, M. 2024, ApJ, 964, 52
\bibitem[Zucker et al.(2006)]{Zucker06} Zucker, D. B., Belokurov, V., Evans, N. W., et al. 2006, ApJL, 643, L103
\end{thebibliography}
\end{document}